\documentclass{article}

\usepackage{setspace}

\usepackage{amssymb}

\usepackage{verbatim,amsmath}
\begin{document}

\title{Bayesian filtering for multi-object systems
with independently generated observations}


\author{Daniel Edward Clark}

\maketitle
\begin{abstract}
A general approach for Bayesian 
filtering of 
multi-object systems is studied, 
with particular emphasis on the model
where each object generates observations independently of other objects.
The approach is based on variational calculus 
applied to generating functionals,
using the 
general version of
Fa\`{a} di Bruno's formula for G\^{a}teaux differentials.
This result enables us to determine some 
general formulae for the updated generating functional
after the application of a multi-object analogue of 
Bayes' rule.

\end{abstract}

%

\section{Introduction}


Many control applications are concerned with the
estimating the state of 
dynamical systems that have uncertainty in their dynamical
and observation characteristics.
Stochastic filtering
methods for discrete-time systems are of particular importance
and have widespread applicability. 
The extension of stochastic filtering to multi-object
systems was
developed for discrete-time systems as
a means of identifying and tracking an unknown 
number of objects in aerospace applications~\cite{Mahlermaths}.
This was based on the Chapman-Kolmogorov equation 
for stochastic population processes proposed by Moyal~\cite{Moyal62},
				and the derivation of Bayes' rule for point processes by Mahler~\cite{Mahlermaths}.
				The mathematics required is based on the theory of generating functionals~\cite{Volterra}
				for describing the evolution of the multi-object system,
								which is closely related to the methods used in quantum field theory~\cite{ZinnJustin}
								for describing the evolution of many-particle systems~\cite{Kiev}.
												The development of stochastic filtering concepts for multi-object systems 
												has received considerable attention in the signal processing literature
												in the last decade for multi-target tracking applications~\cite{VoMaSPS05,Vojournal,BTVO_CPHD_TSP,Mahlermaths,MahlerCPHD,delmoralcaron}. 

This paper describes a general 
functional form for the Bayesian estimator
when each object generates observations independently
of other objects using the higher-order chain rule for G\^{a}teaux differentials~\cite{hocr}.
This result generalises the 
approach for Bayesian estimation 
of multi-object systems proposed by Mahler~\cite{Mahlermaths,MahlerCPHD},
that has been developed for applications in 
multi-target tracking and 
sensor control~\cite{SinghVo,reward,caron}.

The paper is structured as follows.
In the next section, the necessary background on
generating functionals and G\^ateaux differentials
is provided.
In section 3, point processes and conditional
point processes are described with generating functionals.
In section 4, the Chapman-Kolmogorov equation is
briefly described to place the 
work in the context of Bayesian filtering.
The main results of the paper are in
section 5 which describes general Bayesian
estimation for multi-object systems
and derives the general formula 
for the case where objects generate observations
				independently of other objects.

\section{Generating functionals}

The theory of functionals
for describing the state of systems of with variable
numbers of objects
was developed by Volterra
through the concept of generating functionals~\cite{Volterra}.
%
In quantum physics, this idea was 
adopted by Fock~\cite{Fock} 
and Berezin~\cite{Berezin}
for describing an indefinite number of 
particles in the method of second quantization.
In probability theory,
Moyal described stochastic population processes
with the probability generating functional (p.g.fl.)~\cite{Moyal62},
which can be viewed as a generalisation of the 
probability generating function (p.g.f.) to uniquely 
characterise a point process.
Factorial moment densities and Janossy densities
of the point process are found by finding variations of the p.g.fl.
with G\^{a}teaux differentials~\cite{Gateaux}.

This section discusses generating functionals
as a means of describing systems
with a variable number of objects.
 G\^{a}teaux differentials are discussed as a means 
of detemining their constituent functions.
The higher-order chain rule is introduced
to determine the differentials of composite 
functionals~\cite{hocr}.
Finally, the scalar product in Fock space~\cite{Fock}
is used to operate on two functionals. 

\medskip
\noindent
{\bf Definition: G\^{a}teaux differential}
\\
A functional can be viewed as a function that takes
functions as its argument.
A  {\it G\^{a}teaux differential}~\cite[p71-74]{Hille},
of a functional $\Psi(\psi)$
with increment $\xi$, can be defined with
\begin{align}
\delta\Psi(\psi;\xi)
=
\lim_{\epsilon\rightarrow 0}
{1\over\epsilon}
\left(\Psi(\psi+\epsilon\xi)-\Psi(\psi)\right).
\end{align}
The $n^{th}$-order differential (or variation), 
which is defined by recursively,
is denoted $\delta^n\Psi(\psi,\xi_1,\ldots,\xi_n)$.
When this is evaluated at the Dirac delta functions
$\xi_1=\delta_{x_1},\ldots,\xi_n=\delta_{x_n}$,
centred at $x_1,\ldots,x_n$,
we write this as
\begin{align}
{\delta^n
}
\Psi(\psi;
\delta\psi(x_1)\ldots\delta\psi(x_n)
).
\end{align}

\medskip
\noindent
{\bf Definition: generating functional}
\\
Every functional $\Psi(\psi)$ continuous in the field
of continuous functions can be approximated 
by 
a {\it generating functional}
\begin{align}
\label{func}
\Psi(\psi)
=
\zeta_0+
\sum_{n=1}^\infty
{1\over n!}
\left(
\int
\prod_{i=1}^n
dx_i\psi(x_i)
\right)
\zeta_{n}(x_1,\ldots,x_n),
\end{align}
where the functions $\zeta_{n}(x_1,\ldots,x_n)$
are continuous functions independent of the variable $\psi(x)$
that are symmetric in their arguments.
For compactness of notation, we write this as
\begin{align}
\label{func}
\Psi(\psi)
=
\sum_{n=0}^\infty
{1\over n!}
\left(
\int
\prod_{i=1}^n
dx_i\psi(x_i)
\right)
\zeta_{n}(x_1,\ldots,x_n).
\end{align}
This result, due to Volterra~\cite[p20]{Volterra},
generalises Weierstrass'
theorem on continuous functions
represented by polynomials,
to the case of an infinite number of variables.
This series is convergent if $|\psi(x)|<\rho$, 
for some radius of convergence $\rho$.

\noindent
The functions $\zeta_n(x_1,\ldots,x_n)$~\cite{Volterra}, 
can be recovered by finding the G\^{a}teaux differentials with
\begin{align}
\label{evalzero}
\left.{\delta^n
\Psi(\psi;
\delta\psi(x_1)\ldots\delta\psi(x_n)}
)\right|_{\psi=0}=
\zeta_n(x_1,\ldots,x_n).
\end{align}

\medskip
\noindent
{\bf Lemma 1: Fa\`a di Bruno's formula} 
\\
{
Let $\Pi$ be the set of all  partitions of
variables $\eta_1,\ldots,\eta_n$, and
$\pi\in\Pi$ denote a single partition
that has constituent blocks $\omega\in\pi$ of size $|\omega|$
consisting of constituent elements
$\zeta_{\pi,\omega,1},\ldots, \zeta_{\pi,\omega,|\omega|}\in \{\eta_1,\ldots,\eta_n\}$.
The $n^{th}$-order variation of composition $f\circ g$
with increments $\eta_1, \ldots, \eta_n$ is given by
}
\begin{align}
\label{ee}
\delta^n
\left(
\left(f\circ g\right)\left(y\right);
{\eta_1,\ldots,\eta_n}
\right)
=
\sum_{\pi\in \Pi}
\delta^{(|\pi|)}f
\left(
g(y);
\xi_{\pi,1}(y),\ldots,\xi_{\pi,|\pi|}(y)
\right),
\end{align}
{
where $\xi_{\pi,\omega}(y)$
is the variation of order $|\omega|$ with increments
$\zeta_{\pi,\omega,1},\ldots,\zeta_{\pi,\omega,{|\omega|}}$, i.e.}
\begin{align}
\xi_{\pi,\omega}(y)=
\delta^{(|\omega|)}g
\left(
y;
\zeta_{\pi,\omega,1},\ldots,\zeta_{\pi,\omega,{|\omega|}}
\right).
\end{align}
\\
\noindent
{\bf Proof} 
\\
See Clark~\cite{hocr}. 

\medskip
\noindent
{\bf Lemma 2:  Leibniz' rule for functionals} 
\\
Leibniz' rule for functional derivatives 
the differentials of the product
$f(y)g(y)$ with increments
$\Xi=\{\eta_1,\ldots,\eta_n\}$
is
\begin{align}
\delta^n\left(f(y) {g(y)};\eta_1,\ldots,\eta_n\right)
=
\sum_{\Phi\subset \Xi}\delta^{|\Phi|}f\left(y; \phi_1,\ldots,\phi_{|\Phi|}\right)
\delta^{|\Xi|-|\Phi|}g\left(y; \xi_1,\ldots,\xi_{|\Xi|-|\Phi|}\right),
\end{align}
where the
summation is over all subsets $\Phi$ of the increments,
and $\Phi=\{\phi_1,\ldots,\phi_{|\Phi|}\}$, 
$\Xi-\Phi=\{\xi_1,\ldots,\xi_{|\Xi|-|\Phi|}\}$.

\noindent
{\bf Proof}
\\
See Mahler~\cite[p389]{Mahlerbook}.

\bigskip
\noindent
We recall a result from Clark~\cite{hocr},
which shall be used to determine the first-order factorial moment density
of a point process in section 3.

\medskip
\noindent
{\bf Lemma 3}
\\
Let us consider the
can be used for the $n^{th}$ variation
of composite $f(g(y))$
with  increments
$\xi_1(y),\ldots,\xi_n(y)$,
i.e.
\begin{align}
\delta^{(n)}f
\left(
g(y);
\xi_{1}(y),\ldots,\xi_{n}(y)
\right)
\end{align}
Then the differential of this becomes
\begin{align}
\delta
\left(
\delta^{(n)}f
\left(
g(y);
\xi_{1}(y),\ldots,\xi_{n}(y)
\right); \eta
\right)
&=
\delta^{(n+1)}f
\left(
g(y);
\xi_{1}(y),\ldots,\xi_{n}(y),\delta^{1}g\left(y;\eta\right)
\right)
\\&\notag
+
\sum_{\omega=1}^n
\delta^{(n)}f
\left(
g(y);
\xi_{1}(y),\ldots,\delta{\xi}_{\omega}(y;\eta),\dots,\xi_{n}(y)
\right).
\end{align}
\noindent
{\bf Proof} 
\\
See Clark~\cite{hocr}. 

\medskip
\noindent
{\bf Definition: scalar product}
\\
Let us suppose that we have two functionals
$\Psi_1(\psi)$ and $\Psi_2(\psi)$, 
represented by generating functionals, 
so that
\begin{align}
\Psi_1(\psi)&
=\sum_{n=0}^\infty
{1\over n!}
\left(
\int
\prod_{i=1}^n
dx_i\psi(x_i)
\right)
\psi_{1,n}(x_1,\ldots,x_n),
\\
\Psi_2(\psi)&
=\sum_{n=0}^\infty
{1\over n!}
\left(
\int
\prod_{i=1}^n
dx_i\psi(x_i)
\right)
\psi_{2,n}(x_1,\ldots,x_n),
\end{align}
where $\psi_{1,n}(x_1,\ldots,x_n)$ and $\psi_{1,n}(x_1,\ldots,x_n)$
are real-valued.
We can define a {\it scalar product}
of $\Psi_1(\psi)$ and $\Psi_2(\psi)$ is 
 with~\cite[p164]{ZinnJustin}
\begin{align}
\label{scalar}
\left\langle
\Psi_1,\Psi_2
\right\rangle
&=
\sum_{n=0}^\infty
{1\over n!}
\left(
\int
\prod_{i=1}^n
dx_i
\right)
{\psi}_{1,n}(x_1,\ldots,x_n)
\psi_{2,n}(x_1,\ldots,x_n).
\end{align}

\section{Probability generating functionals}
\noindent
Following Moyal~\cite{Moyal62}, define the {\it probability generating functional}
(p.g.fl.) 
with
\begin{align}
    G_X(\psi)&
    =\sum_{n=0}^\infty
    {1\over n!}
    \left(
    \int
    \prod_{i=1}^n
    dx_i\psi(x_i)
    \right)
    p_{n}(x_1,\ldots,x_n),
    \end{align}
where $p_{n}(x_1,\ldots,x_n)\ge0$ are Janossy densities and
$G_X(\psi)=1$.
We refer to the sequence of functions $p_{n}(x_1,\ldots,x_n)$ 
as a {\it multi-object probability density}.
The p.g.fl. is commonly used in point process theory
to characterise point patterns~\cite{RCoxIsham}.
The Janossy densities $p_n(x_1,\ldots,x_n)$
and factorial moment densities $M_n(x_1,\ldots,x_n)$
can be recovered from the p.g.fl. by finding its
variations~\cite{Volterra, Moyal62}, i.e.
\begin{align}
p_n(x_1,\ldots,x_n)&=
\left.
{\delta^n
G_X(\psi;
\delta\psi(x_1)\ldots\delta\psi(x_n) }
)\right|_{\psi=0},
\\
M_n(x_1,\ldots,x_n)&=\left.
{\delta^n
G_X(\psi;
\delta\psi(x_1)\ldots\delta\psi(x_n) }
)\right|_{\psi\rightarrow 1}
.\end{align}

\medskip
\noindent
{\bf Definition: conditional probability
    generating functional}
\\
Let us define a 
generating functional
$G_{Z|X}(\psi|\eta,\phi)$
of the form
\begin{align}
\label{cpg}
G_{Z|X}(\psi|\eta,\phi)
&=
\sum_{n=0}^\infty
{1\over n!}
\left(\int\prod_{j=1}^ndx_j\eta(x_j)\phi(x_j)\right)
H_{Z|X}(\psi|x_1,\ldots,x_n)
\end{align}
where 
\begin{align}
\label{pgfllik}
H_{Z|X}(\psi|x_1,\ldots,x_n)
&=
\sum_{m=0}^\infty
{1\over m!}
\left(\int\prod_{i=1}^mdz_i\psi(z_i)\right)
p_{m|n}(z_1,\ldots,z_m|x_1,\ldots,x_n),
\end{align}
and for any $n\ge0$ and $\Xi=\{x_1,\ldots,x_n\}$, 
$p_{m|n}(z_1,\ldots,z_m|x_1,\ldots,x_n)\ge0$,
\begin{align}
    \sum_{m=0}^\infty
    {1\over m!}
 \left(\int\prod_{i=1}^mdz_i\right)
    p_{m|n}(z_1,\ldots,z_m|x_1,\ldots,x_n)=1.
\end{align}
The sequence $p_{m|n}(z_1,\ldots,z_m|x_1,\ldots,x_n)$
defines a {\it conditional multi-object probability density},
and $G_{Z|X}(\psi|\eta,\phi)$ defines a 
{\it conditional probability
    generating functional.}

\medskip
\noindent
{\bf Theorem 1}
\\
Let us define
\begin{align}
&q_{k|m}(x_1,\ldots,x_k|z_1,\ldots,z_m) 
=
\\\notag&
{
 p_{m|k}(z_1,\ldots,z_m|x_1,\ldots,x_k) p_{k}(x_1,\ldots,x_k) 
\over
\sum_{n=0}^\infty
{1\over n!}
\left(\int\prod_{j=1}^ndy_j \right)
p_{m|n}(z_1,\ldots,z_m|y_1,\ldots,y_n)
p_{n}(y_1,\ldots,y_n)
}.\label{bu}
\end{align}
Then 
the sequence of functions
$q_{k|m}(x_1,\ldots,x_k|z_1,\ldots,z_m)$
is a conditional multi-object probability density.

\medskip
\noindent
{\bf Proof }
\\
For any set $Z=\{z_1,\ldots,z_m\}$,
we can evaluate the infinite sum
\begin{align}
&\sum_{k=0}^\infty
{1\over k!}
\left(\int\prod_{j=1}^kdy_j \right)
q_{k|m}(y_1,\ldots,y_k|z_1,\ldots,z_m).
\end{align}
Substituting this into equation (\ref{bu}), we
get
\begin{align}
{\sum_{k=0}^\infty
{1\over k!}
\left(\int\prod_{j=1}^kdy_j \right)
     p_{m|k}(z_1,\ldots,z_m|x_1,\ldots,x_k) p_{k}(x_1,\ldots,x_k) \over
\sum_{n=0}^\infty
{1\over n!}
\left(\int\prod_{j=1}^ndy_j \right)
p_{m|n}(z_1,\ldots,z_m|y_1,\ldots,y_n)
p_{n}(y_1,\ldots,y_n)
}=1.
\end{align}
Thus
the sequence of functions
$q_{k|m}(x_1,\ldots,x_k|z_1,\ldots,z_m)$
defines a conditional multi-object probability density. 

\medskip
\noindent
{\bf Corollary} 
\\
The 
conditional multi-object density function
defined with sequence of functions
$q_{k|m}(y_1,\ldots,y_k|z_1,\ldots,z_m)$
in Theorem 1
is a multi-object analogue of Bayes' rule.
Thus, if we have a set of
measurements $Z=\{z_1,\ldots,z_m\}$,
and multi-object prior probability density,
we can determine a conditional posterior multi-object probability
density.
\bigskip

\noindent

\section{The Chapman-Kolmogorov equation}

				For a single-object system, the Chapman-Kolmogorov equation
				propagates a posterior density $p_k(x')$ in state $x'$ at time-step $k$,
				with a Markov transition $f_M(x|x')$ to time-step $k+1$ 
				to give predicted density $p_{k+1|k}(x)=\int f_M(x|x')p_k(x')dx'$.
				In multi-object systems, we need to consider all joint densities, 
				which motivates the use of probability generating functionals.
				Let us define a conditional probability generating functional
				$G_M(\psi|\eta)$
				that describes the Markov transition of the multi-object system,
				\begin{align}
				&G_M(\psi|\eta)
				=
				\\\notag
				&\sum_{n=0}^\infty
				\sum_{m=0}^\infty
				{1\over n!}
				{1\over m!}
				\left(\int \prod_{i=1}^ndx_i\psi(x_i)\right)
				\left(\int \prod_{j=1}^mdy_j\eta(y_j)\right)
				p_{M,n|m}(x_1,\ldots,x_n|y_1,\ldots,y_m),
				\end{align}
				where
				$p_{M,n|m}(x_1,\ldots,x_n|y_1,\ldots,y_m)$
				is the probability of transition to the multi-object state $x_1,\ldots,x_n$
				at the next time-step
				from multi-object state $y_1,\ldots,y_m$ at the current time-step.
				Then, using the scalar product for functionals in Fock space, the predicted p.g.fl.~\cite{Moyal62,Mahlermaths} is given by
				\begin{align}
				G_{k+1|k}(\psi)=\langle G_M(\psi|\cdot), G_k\rangle,
				\end{align}
				where $G_k$ is the p.g.fl. for the posterior probability 
				density at time-step $k$.
				This follows straightforwardly from the definition of the scalar
				product.
The Janossy densities and factorial moment densities
can be determined via G\^{a}teaux differentials.
Since we are primarily focussed on Bayesian estimation,
we do not provide further details.

\section{Bayesian estimation}
For a single-object system, Bayes' rule updates a prior $p(x)$
with measurement $z$ using likelihood $g(z|x)$
using the formula $p(x|z)=p(x)g(z|x)/\int p(y) g(z|y)dy$.
This section develops Bayesian estimation for
multi-object systems
based on Mahler's approach
for determining the posterior p.g.fl.
based on a set of measurements~\cite{Mahlermaths}.
Section 3.1 derives the general multi-object Bayesian 
estimator.
Section 3.2 describes the Bayesian estimator
where each object generates observations
independently of other objects 
and derives the general generating functional form.

\subsection{General multi-object Bayesian estimator}

The probability generating functional
and conditional probability generating functional
are defined.
Theorem 1 derives a multi-object analogue of
Bayes' rule.
Lemma 4 provides a means of determining the
updated multi-object probability density via 
generating functionals.
This was originally shown by Mahler~\cite{Mahlermaths}
using a bivariate functional 
\begin{align}
F[\psi,\eta]=\langle G_{Z|X}(\psi|\eta,\cdot),G_X\rangle.
\end{align}


\noindent
{\bf Lemma 4}
\\
Consider the 
scalar product $\langle G_{Z|X}(\psi|\eta,\cdot),G_X\rangle$
of the form (\ref{cpg}) and prior 
p.g.fl. $G_X$,
and let us define the conditional
generating functional
\begin{align}
\label{thegeneralpgfl}
G_{X|Z}(\eta|z_1,\ldots,z_m)=
{\left.
\delta^m
\left({\langle G_{Z|X}(\psi|\eta,\cdot),G_X\rangle;
\delta\psi(z_1)\ldots\delta\psi(z_m)}\right)
\right|_{\psi=0}
\over
\left.
\delta^m
\left({
\langle G_{Z|X}(\psi|1,\cdot),G_X\rangle
;\delta\psi(z_1)\ldots\delta\psi(z_m)}
\right)
\right|_{\psi=0}
}.
\end{align}
\noindent
The sequence of functions $q_{k|m}(x_1,\ldots,x_k|z_1,\ldots,z_m)$
can be recovered 
by finding the variations as follows
\begin{align}
q_{k|m}(x_1,\ldots,x_n|z_1,\ldots,z_m)&=
\left.
{\delta^k
G_{X|Z}(\eta|z_1,\ldots,z_m;\delta\eta(x_1)\ldots\delta\eta(x_k)}
\right|_{\eta=0}.
\end{align}

\noindent
{\bf Proof}
\\
The scalar product
evaluates to be
\begin{align}
&\langle G_{Z|X}(\psi|\eta,\cdot),G_X\rangle
=
\sum_{n=0}^\infty
{1\over n!}
\left(\int\prod_{j=1}^ndx_j\eta(x_j)\right)
H_{Z|X}(\psi|x_1,\ldots,x_n)
p_n(x_1,\ldots,x_n).
\end{align}
Finding the variations of this leads to
\begin{align}
&
\left.
\delta^m
\left(
{\langle G_{Z|X}(\psi|\eta,\cdot),G_X\rangle
;\delta\psi(z_1)\ldots\delta\psi(z_m)}\right)
\right|_{\psi=0}
\\\notag
\\\notag&
=
\sum_{n=0}^\infty
{1\over n!}
\left(\int\prod_{j=1}^ndx_j\eta(x_j)\right)
p_{m|n}(z_1,\ldots,z_m|x_1,\ldots,x_n)
p_n(x_1,\ldots,x_n).
\end{align}

\noindent
Hence, if we subsequently find the variations of
$G_{X|Z}(\eta|z_1,\ldots,z_m)$, we get
\begin{align}
&\left.
{\delta^k
G_{X|Z}(\eta|z_1,\ldots,z_m;
\delta\eta(x_1)\ldots\delta\eta(x_k)}
)
\right|_{\eta=0}
\\\notag
\\\notag
&=
{
\left.
\delta^{k+m}
\left(
\langle G_{Z|X}(\psi|\eta,\cdot),G_X\rangle
;\delta\eta(x_1)\ldots\delta\eta(x_k),
\delta\psi(z_1)\ldots\delta\psi(z_m)
\right)\right|_{\psi=0,\eta=0}
\over
\left.
\delta^m
\left(\langle G_{Z|X}(\psi|1,\cdot),G_X\rangle
;\delta\psi(z_1)\ldots\delta\psi(z_m)\right)
\right|_{\psi=0}
}
\\\notag
\\\notag
&={
p_{m|k}(z_1,\ldots,z_m|x_1,\ldots,x_k)
p_k(x_1,\ldots,x_k)
\over
\sum_{n=0}^\infty
{1\over n!}
\left(\int\prod_{j=1}^ndx_j\right)
p_{m|n}(z_1,\ldots,z_m|x_1,\ldots,x_n)
p_n(x_1,\ldots,x_n)}
\\\notag
\\\notag
&=
q_{k|m}(x_1,\ldots,x_k|z_1,\ldots,z_m).
\end{align}

\medskip

\noindent
{\bf Corollary}
\\
The conditional generating functional
$G_{X|Z}(\eta|z_1,\ldots,z_m)$
defines the probability generating functional
for determining the Janossy densities 
$q_{k|m}(x_1,\ldots,x_k|z_1,\ldots,z_m)$.
We refer to $G_{X|Z}(\eta|z_1,\ldots,z_m)$
as the {\it p.g.fl. Bayes update}.

\bigskip

\noindent
In the next section we use the general p.g.fl. Bayes update to determined the 
p.g.fl. for the case where each object generates independently of other objects.

\subsection{Independently generated observations}

This section considers the scenario where 
the p.g.fl. likelihood defined in
(\ref{pgfllik}) becomes
\begin{align}
H_{Z|X}(\psi|x_1,\ldots,x_n)=
\prod_{i=1}^n
G_{Z|x}(\psi|x_i),
\end{align}
where
\begin{align}
\label{wwq}
G_{Z|x}(\psi|x)=
\sum_{m=0}^\infty
{1\over m!}
\int\prod_{i=1}^mdz_i\psi(z_i)
r_{m|1}(z_1,\ldots,z_m|x),
\end{align}
and $r_{m|1}(z_1,\ldots,z_m|x)$
is a Janossy density conditioned on a single object $x$.
Moyal defined this named this process
a {\it multiplicative population process}~\cite{Moyal62}
and showed that it represents the scenario where 
objects generated measurements 
independently of other objects.
In this section we provide the general formula
for this model in terms of its p.g.fl..
Lemma 5 describes the scalar product from section 3.1,
which is used in Theorem 2 to derive the generating
functional form of Bayes' rule.
This is of the form of a composite of functionals,
which enables us to use the higher-order chain for G\^{a}teaux differentials~\cite{hocr}.
The resulting formula is demonstrated with 
the example of a Poisson point process.
The general first-order moment is then derived for this model.

\medskip
\noindent{\bf Lemma 5}\\
The scalar product defined in (\ref{cpg}) 
for this model
becomes
\begin{align}
\langle G_{Z|X}(\psi|\eta,\cdot),G_X\rangle
&=
G_X\left(\eta G_{Z|x}(\psi|\cdot)\right).
\end{align}

\medskip
\noindent{\bf Proof}\\
The conditional probability generating functional
$G_{Z|X}(\psi|\eta,\phi)$ in equation (\ref{cpg}) becomes
\begin{align}
G_{Z|X}(\psi|\eta,\phi)
&=
\sum_{n=0}^\infty
{1\over n!}
\int\prod_{j=1}^ndx_j\eta(x_j)\phi(x_j)
G_{Z|x}(\psi|x_j)
.
\end{align}
Hence the 
scalar product becomes
\begin{align}
\langle G_{Z|X}(\psi|\eta,\cdot),G_X\rangle
&=
\sum_{n=0}^\infty
{1\over n!}
\left(\int\prod_{j=1}^ndx_j\eta(x_j)
G_{Z|x}(\psi|x_j)
\right)
p_n(x_1,\ldots,x_n)
\\\notag
&=
G_X\left(\eta G_{Z|x}(\psi|\cdot)\right),
\end{align}
as required. 

\bigskip

\noindent
For compactness of notation in Theorem 2, 
let us introduce the notation
$P_{0}(x)=r_{0|1}(x)$,
$P_{|Z|}(Z|x)= r_{m | 1}(z_1,\ldots,z_m|x)$,
where $Z=\{z_1,\ldots,z_m\}$.

\medskip
\noindent
{\bf Theorem 2}
\\
Let us suppose that $\Pi$ is the set of all partitions of the 
measurement set $Z$.
The general p.g.fl. Bayes update for this model becomes
\begin{align}
\label{generalpgfl}
G_{X|Z}(\eta|z_1,\ldots,z_m)
=
{\sum_{\pi\in \Pi}
\delta^{|\pi|}
G_X
\left(
\eta
P_{0};
\eta 
P_{|Z_{\pi,1}|}(Z_{\pi,1}|\cdot)
,\ldots,
\eta
P_{|Z_{\pi,|\pi|}|}(Z_{\pi,|\pi|}|\cdot)
\right)
\over
\sum_{\pi\in \Pi}
\delta^{|\pi|}
G_X
\left(
P_{0};
P_{|Z_{\pi,1}|}(Z_{\pi,1}|\cdot)
,\ldots,
P_{|Z_{\pi,|\pi|}|}(Z_{\pi,|\pi|}|\cdot)
\right)
}
,
\end{align}
where for each partition $\pi$,
$Z$ is the disjoint union of subsets $Z_{\pi,j}\subset Z$.

\medskip
\noindent
{\bf Proof}
\\
The $m^{th}$-order variation of the scalar product 
$\langle G_{Z|X}(\psi|\eta,\cdot),G_X\rangle$
simplifies
to finding the $m^{th}$-order variation 
of
composite $G_X\left(\eta G_{Z|x}(\psi|\cdot)\right)$, so that

\begin{align}
\label{kth}
&
\delta^m\left(
\langle G_{Z|X}(\psi|\eta,\cdot),G_X\rangle;
\delta\eta(z_1),\ldots,\delta\eta(z_m)\right)
=
\delta^m\left(
G_X\left(\eta G_{Z|x}(\psi|\cdot)\right);
\delta\eta(z_1),\ldots,\delta\eta(z_m)\right).
\end{align}

\medskip
\noindent Using the higher-order chain rule, 
we find that 
this is equal to
\begin{align}
\label{tte}
\sum_{\pi\in \Pi}
\delta^{|\pi|}
G_X
\left(
\eta
G_{Z|x}(\psi|\cdot);
\eta
\alpha_{\pi,1}(\psi|\cdot) ,\ldots,
\eta
\alpha_{\pi,|\pi|}(\psi|\cdot) 
\right),
\end{align}
where the sum is over all partitions of measurement set $Z$,
The increments $\alpha_{\pi,\omega}(\psi|\cdot)$
are found with the variations of $G_{Z|x}(\psi|\cdot)$, where
\begin{align}
\label{alpha}
\alpha_{\pi,\omega}(\psi|\cdot)=
\delta^{|\omega|}G_{Z|x}
\left(\psi|\cdot;
\zeta_{\pi,\omega,1}
,\ldots,
\zeta_{\pi,\omega,{|\omega|}}
\right),
\end{align}
and each $\zeta_{\pi,\omega,i}\in\{\delta\eta(z_{1}),\ldots,\delta\eta(z_{m})\}$ is counted once for each partition $\pi$ of the increments.
Then setting $\psi=0$, we have
\begin{align}
G_{Z|x}(0|x)
&=P_{0}\left(x\right),
\\
\alpha_{\pi,\omega}(0|x)&=
P_{|Z_{\pi,\omega|}|}\left(Z_{\pi,\omega}|x\right).
\end{align}
Substituting this into (\ref{tte}) and 
(\ref{thegeneralpgfl})
yields the result. 

\medskip
\noindent{\bf Corollary}
\\
The result given above shows that the p.g.fl. Bayes update can be described in terms of variations of the prior p.g.fl..
Since the p.g.fl. uniquely characterises a point process,
it suffices to present the updated point processes in terms of their p.g.fl..
This provides a general formula for the model 
introduced by Mahler~\cite{Mahlermaths,MahlerCPHD}.

\medskip
\noindent{\bf Example: Poisson point process}
\\
This parameterisation was considered in the Probability Hypothesis Density (PHD) filter proposed by Mahler~\cite{Mahlermaths}.
Let us suppose that we have a Poisson point process with intensity $\mu(x)$.
The p.g.fl. becomes~\cite[p167]{DaleyVere-Jones2v1}
\begin{align}
G_X(\eta)=\exp\left(\mu\left[\eta-1\right]\right)
\label{poissonp}
\end{align}
where we adopt the functional notation $\mu[\eta]=\int \mu(x)\eta(x)dx$.
Then the p.g.fl. Bayes update is
\begin{align}
G_{X|Z}(\eta|z_1,\ldots,z_m)&=
\exp\left(\mu\left[ (\eta-1)P_{0}\right]\right) 
{
\sum_{\pi\in\Pi}
\prod_{i=1}^{|\pi|} \mu\left[ \eta P_{|Z_{\pi,i}|}(Z_{\pi,i} |\cdot)\right]
\over
\sum_{\pi\in\Pi}
\prod_{i=1}^{|\pi|} \mu\left[   P_{|Z_{\pi,i}|}(Z_{\pi,i} |\cdot)\right]
}.
\end{align}

\medskip
\noindent{\bf Proof}
\\
The $k^{th}$-order variation of the Poisson p.g.f.l. with increments $\eta_1,\ldots,\eta_k$ is 
\begin{align}
\delta^kG_X\left(\eta;\eta_1,\ldots,\eta_k\right)
=
\exp\left(\mu\left[\eta-1\right]\right)  \prod_{i=1}^k \mu \left[ \eta_i \right]
.\label{poissond}
\end{align}
The result follows by substitution into equation (\ref{generalpgfl}). 

\bigskip 
\noindent The first-order factorial moment density, 
 known as the intensity function in the point process literature,
is typically used in signal processing applications~\cite{Mahlermaths},
which can be found by further G\^{a}teaux differentiation.
We now use Theorem 2 to find the first-order factorial moment density in Theorem 3.

\medskip
\noindent
{\bf Theorem 3}
\\
The general form of the first-order factorial moment density 
$M_1(x)$
is given by
\begin{align}
\label{general1stp}
M_1(x)&=
{\sum_{\pi\in \Pi}
\delta^{|\pi|+1}
G_X
\left(
P_{0};
P_{|Z_{\pi,1}|}(Z_{\pi,1}|\cdot)
,\ldots,
P_{|Z_{\pi,|\pi|}|}(Z_{\pi,|\pi|}|\cdot),
P_{0}(x)
\right)
\over
\sum_{\pi\in \Pi}
\delta^{|\pi|}
G_X
\left(
P_{0};
P_{|Z_{\pi,1}|}(Z_{\pi,1}|\cdot)
,\ldots,
P_{|Z_{\pi,|\pi|}|}(Z_{\pi,|\pi|}|\cdot)
\right)
}
\\\notag
\\\notag
&+
{\sum_{\pi\in \Pi}
\sum_{i=1}^{|\pi|}
\delta^{|\pi|}
G_X
\left(
P_{0};
P_{|Z_{\pi,1}|}(Z_{\pi,1}|\cdot)
,\ldots,
 P_{|Z_{\pi,i}|}(Z_{\pi,i}|x),
\ldots,
 P_{|Z_{\pi,|\pi|}|}(Z_{\pi,|\pi|}|\cdot)
\right)
\over
\sum_{\pi\in \Pi}
\delta^{|\pi|}
G_X
\left(
P_{0};
P_{|Z_{\pi,1}|}(Z_{\pi,1}|\cdot)
,\ldots,P_{|Z_{\pi,|\pi|}|}(Z_{\pi,|\pi|}|\cdot)
\right)
},
\end{align}
where the $i^{th}$ increment in the second summation
is evaluated at $P_{|Z_{\pi,i}|}(Z_{\pi,i}|x)$.

\medskip
\noindent
{\bf Proof}
\\
Using the product rule for G\^{a}teaux differentials
from section 2,
we get
\begin{align}
\label{general1st}
&{\delta} G_{X|Z}(\eta;\xi|z_1,\ldots,z_m)
=
{\sum_{\pi\in \Pi}
\delta^{|\pi|+1}
G_X
\left(
\eta
P_{0};
\eta 
P_{|Z_{\pi,1}|}(Z_{\pi,1}|\cdot)
,\ldots,
\eta
P_{|Z_{\pi,|\pi|}|}(Z_{\pi,|\pi|}|\cdot),
\xi P_{0}
\right)
\over
\sum_{\pi\in \Pi}
\delta^{|\pi|}
G_X
\left(
P_{0};
P_{|Z_{\pi,1}|}(Z_{\pi,1}|\cdot)
,\ldots,
P_{|Z_{\pi,|\pi|}|}(Z_{\pi,|\pi|}|\cdot)
\right)
}
\\\notag
\\\notag
&+
{\sum_{\pi\in \Pi}
\sum_{i=1}^{|\pi|}
\delta^{|\pi|}
G_X
\left(
\eta
P_{0};
\eta 
P_{|Z_{\pi,1}|}(Z_{\pi,1}|\cdot)
,\ldots,
\xi P_{|Z_{\pi,i}|}(Z_{\pi,i}|\cdot),
\ldots,
\eta P_{|Z_{\pi,|\pi|}|}(Z_{\pi,|\pi|}|\cdot)
\right)
\over
\sum_{\pi\in \Pi}
\delta^{|\pi|}
G_X
\left(
P_{0};
P_{|Z_{\pi,1}|}(Z_{\pi,1}|\cdot)
,\ldots,P_{|Z_{\pi,|\pi|}|}(Z_{\pi,|\pi|}|\cdot)
\right)
}
,
\end{align}
where $\xi$ replaces $\eta$ in the $i^{th}$ increment,
since when $g(\eta)=\eta P$,
\begin{align}
\delta g (\eta ;\xi)= \xi P,
\end{align}
which follows directly from the definition of the G\^{a}teaux differential.
Letting $\eta\rightarrow 1$ and $\xi=\delta_x$, 
we get the desired result.


\bigskip
\noindent
{\bf Corollary}\\
Theorem 3 allows us to determine the updated intensity function
of a point process after the application of the multi-object analogue of Bayes'
rule.
This provides a general result for 
multi-object filters
where it assumed that objects generate measurements independently of
 each other, introduced by Mahler~\cite{Mahlermaths}.

\bigskip
\noindent
{\bf Example: Poisson point process}\\
Following the Poisson point process described earlier,
the first-order factorial moment density $M_1(x)$ is given by
\begin{align}
\label{general1stpoisson}
M_1(x)&=
{\mu(x)
}
\left(
{\sum_{\pi\in \Pi}
\prod_{i=1}^{|\pi|} \mu\left[  P_{|Z_{\pi,i}|}(Z_{\pi,i} |\cdot)\right]
\left(
P_0(x)
+
\sum_{j=1}^{|\pi|}
{
 P_{|Z_{\pi,i}|}(Z_{\pi,i} |x)
\over
\mu\left[ P_{|Z_{\pi,j}|}(Z_{\pi,j} |\cdot)\right]
}\right)
\over
\sum_{\pi\in \Pi} \prod_{i=1}^{|\pi|} \mu\left[  P_{|Z_{\pi,i}|}(Z_{\pi,i} |\cdot)\right]
}
\right).
\end{align}

\medskip
\noindent
{\bf Proof}
\\
The result is an application of Theorem 3 that 
follows by substitution
of  (\ref{poissond}) into equation (\ref{general1st}).

\subsection{Independently generated measurements and uncorrelated observations}

This section extends
the scenario considered in the previous section
by including observations uncorrelated with objects,
so that
\begin{align}
H_{Z|X}(\psi|x_1,\ldots,x_n)=G_\kappa(\psi)
\prod_{i=1}^n
G_{Z|x}(\psi|x_i),
\end{align}
where the generating functional $G_\kappa(\psi)$
is uncorrelated with the objects.
This is the scenario typically used in target 
tracking applications~\cite{Mahlermaths,MahlerCPHD}.
In this section we provide the general formula
for this model in terms of its p.g.fl..
Lemma 6 describes the scalar product from section 3.1,
which is used in Theorem 2 to derive the generating
functional form of Bayes' rule.
This is of the form of a composite of functionals,
which enables us to use the higher-order chain for G\^{a}teaux differentials~\cite{hocr}.
The resulting formula is demonstrated with 
the example of a Poisson point process.
The general first-order moment is then derived for this model.

\medskip
\noindent{\bf Lemma 6}\\
Let us consider
conditional probability generating functional
in equation (\ref{wwq}) 
and 
\begin{align}
\label{wwqf}
G_{\kappa}(\psi)=
\sum_{m=0}^\infty
{1\over m!} \left(\int\prod_{i=1}^mdz_i\psi(z_i)\right)
p_{m|0}(z_1,\ldots,z_m)
\end{align}
Then the 
scalar product
$\langle G_{Z|X}(\psi|\eta,\cdot),G_X\rangle$ becomes
\begin{align}
\langle G_{Z|X}(\psi|\eta,\cdot),G_X\rangle
&=G_{\kappa}(\psi) G_X\left(\eta G_{Z|x}(\psi|\cdot)\right).
\end{align}

\medskip
\noindent{\bf Proof}\\
Since the $G_{\kappa}(\psi)$ functionals
and $G_X\left(\eta G_{Z|x}(\psi|\cdot)\right)$ are,
by definition, independent
we can use the multiplicative property of Moyal~\cite{Moyal62},
which states that the superposition
of two independent point processes 
can be determined through the multiplication of
their p.g.fl.s.

\medskip
\noindent
{\bf Theorem 4}\\
The general p.g.fl. Bayes update for this model becomes
Let us write $p_{m|0}(z_1,\ldots,z_m)=P_\kappa(Z)$,
where $Z=\{z_1,\ldots,z_m)$.

\begin{align}
\label{generalpgfl2}
&
G_{X|Z}(\eta|z_1,\ldots,z_m)
=
\\\notag&
{\sum_{W \subset Z}
P_{\kappa}
\left(
Z \setminus W
\right)
\sum_{\pi\in \Pi_{W}}
\delta^{(|\pi|)}
G_X
\left(
\eta
P_{Z|x}\left(\emptyset|\cdot\right)
;
\eta
P_{Z|x}\left(Z_{\pi,1}|\cdot\right)
,\ldots,
\eta
P_{Z|x}\left(Z_{\pi,|\pi|}|\cdot\right)
\right)
\over
\sum_{W \subset Z}
P_{\kappa}
\left(
Z \setminus W
\right)
\sum_{\pi\in \Pi_{W}}
\delta^{(|\pi|)}
G_X
\left(
P_{Z|x}\left(\emptyset|\cdot\right)
;
P_{Z|x}\left(Z_{\pi,1}|\cdot\right)
,\ldots,
P_{Z|x}\left(Z_{\pi,|\pi|}|\cdot\right)
\right)
}
,
\end{align}
where the left summation is over all subsets $W$ 
of measurement set $Z$,
and the inner summation is over all partitions
of $W$.

\noindent
{\bf Proof}\\
From Leibniz' rule
and equation (\ref{tte}),
the $k^{th}$-order variation of product
$G_{\kappa}(\psi) G_X\left(\eta G_{Z|x}(\psi|\cdot)\right)$
with respect to $\psi$
and increments $\Phi=\{\eta_1,\ldots,\eta_k\}$
becomes
\begin{align}
&\sum_{\Xi\subset\Phi}
\delta^{|\Xi|}
G_{\kappa} \left(\eta ;\zeta_1,\ldots,\zeta_{|\Xi|} \right)
\sum_{\pi\in \Phi-\Xi}
\delta^{|\pi|}
G_X
\left(
\eta
G_{Z|x}(\psi|\cdot);
\eta
\alpha_{\pi,1}(\psi|\cdot) ,\ldots,
\eta
\alpha_{\pi,|\pi|}(\psi|\cdot) 
\right),
\end{align}
where $\alpha_{\pi,j}$
is defined in equation $\ref{alpha}$.
In a similar manner to 
Theorem 2, when we substitute into the
general p.g.fl. in equation 
(\ref{thegeneralpgfl})
we get the required result.

\bigskip

\noindent
{\bf Corollary}\\
The general form of the first-order factorial moment density 
follows from Theorems 3 and 4.
We omit this result for brevity.

\section{Discussion}
This paper provides a general means of Bayesian
estimation for multi-object systems
where each object generates measurements
independently.
Using the higher-order chain rule for
G\^{a}teaux differentials,
a general functional form for Bayesian estimation.
In a forthcoming paper, 
this result will be applied to a range of
point process models for
statistical signal processing applications.

\bibliography{bib_new1}
\end{document}